\documentclass[iop]{emulateapj}
\usepackage{amsmath,amsfonts,amssymb,mathrsfs,graphicx}
\usepackage[squaren]{SIunits}
\usepackage{natbib}

\usepackage{hyperref}

\begin{document}
\title{Constraints on The Hadronic Content of Gamma Ray Bursts}

\author{Lee Yacobi}

\affiliation{Department of Physics, Technion, Israel}

\author{Dafne Guetta}
\affiliation{Osservatorio astronomico di Roma, v. Frascati 33, 00040 Monte Porzio Catone, Italy}
\affiliation{Department of Physics
 Optical Engineering, ORT Braude, P.O. Box 78, Carmiel, Israel}
\affiliation{Department of Physics, Technion, Israel}

\author{Ehud Behar}
\affiliation{Department of Physics, Technion, Israel}

\begin{abstract}
The IceCube high-energy neutrino telescope has been collecting data since 2006.
Conversely, hundreds of Gamma Ray Bursts (GRBs) have been detected by the GBM on board Fermi, since its launch in 2008.
So far no neutrino event has been associated with a GRB,
despite many models predicting the generation of high energy neutrinos through GRB photon interaction with PeV protons in the GRB jet.
We use the non-detection of neutrinos to constrain the hadronic content of GRB jets independent of jet model parameters.
Assuming a generic particle spectrum of $E^{-\alpha}$ with $\alpha = 2$,
we find that the ratio of the energy carried by pions to that in electrons has to be small $f_\pi / f_e \lesssim 0.24$
at 95\% confidence level.
A distribution of spectral slopes can lower $f_\pi / f_e$ by orders of magnitude.
Another limit, independent of neutrinos,
is obtained if one ascribes the measured Fermi/LAT GeV gamma-ray emission to pair-photon cascades of high-energy photons resulting from (the same photon-hadronic interactions and subsequent) neutral pion decays.
Based on the generally observed MeV to GeV GRB fluence ratio of $\approx 10$, we show that $f_\pi / f_e \lesssim 0.3$.
In some bursts, where this ratio is as low as unity,  $f_\pi / f_e \lesssim 0.03$.
These findings add to the mounting doubts regarding the presence of PeV protons in GRB jets.

\end{abstract}

\maketitle

\flushbottom

$~$

$~$
\section{Introduction}
Gamma Ray Bursts (GRBs) are powerful explosions, and are among the highest-redshift point sources observed. The most common phenomenological interpretation of these cosmological sources is through the so called fireball model \citep{fireball:2000, meszaAFTERGLOW, Meszaros06}. In this model, part of the energy is carried out (e.g., from a collapsed star) by hadrons at highly-relativistic energies, some of which
is dissipated internally and eventually
radiated as $\gamma$-rays by synchrotron and inverse-Compton emission by shock-accelerated electrons.
As the fireball sweeps up ambient material, it energizes the surrounding medium through, e.g., forward shocks, which are believed to be responsible for the longer-wavelength afterglow emission \citep{meszaAFTERGLOW}.

If the GRB jet comprises PeV protons, it should produce energetic neutrinos through photon-hadron interactions.
The photons for this process can be supplied by the GRB gamma rays during its prompt phase, or during the afterglow phase \citep{WB97, Dermer02}.
These would lead to the production of charged pions, which subsequently decay to produce neutrinos.
Within this picture, GRBs should produce neutrinos with energies of  $\sim 100$ TeV (observed frame) from the
same region in which the GRB photons are produced \citep{dafneRATE}.
These neutrinos, if present, could be readily detected.
Hence, the detectability of TeV to PeV neutrinos depends on the presence of ($>$) PeV protons and on the efficiency at which their energy is converted into neutrinos, as compared to how much of the energy is in electrons, which is manifested primarily in the prompt GRB photon emission.

The high-energy neutrinos from GRBs should be detected by large neutrino telescopes, such as IceCube and in the future KM3NeT \footnote{http://km3net.org/home.php}. IceCube, is a Cherenkov detector \citep{halzen2010} with photomultipliers (PMTs) at depths between 1450 and 2450 meters in the Antarctic ice designed specifically to detect neutrinos at TeV-PeV energies. Since May 2011 \citep{Aartsen_Apj}, IceCube has been working with a full capacity of 86 strings (IC86).
Since GRB neutrino events need to be correlated both in time and in direction with the gamma-rays, they are sought after in small angular and short time windows.
In this context, IceCube has recently developed a powerful model-independent analysis tool for neutrinos detection, which is coincident in direction and in time to within 1,000 seconds with GRB flares reported by the gamma ray satellites.
IceCube reported no detection of any  GRB-associated neutrino in a data set taken from April 2008 to May 2010  \citep{icecubenature}; None of the high energy neutrinos reported for the next two years \citep{Aartsen_Sci} is GRB-associated either, and as far as we know no neutrino event has been associated with any GRB to date.
This non-detection is in conflict with earlier models \citep{WB97, meszabig, dafnebig, otherbig, he2012},
all of which predicted the detection of approximately ten GRB neutrinos by IceCube during this period.
Those earlier estimates were largely calibrated based on the fireball hypothesis, and were motivated by the assumption that UHECRs are produced primarily by GRBs.
The IceCube results thus appear to rule out GRBs as  the main sources of UHECRs \citep{otherbig, icecubenature}.
This implies either that GRBs do not have the  ($>$)PeV protons, hypothesized in the fireball model,  or that the efficiency of neutrino production from these protons is much lower than had been estimated \citep{small1,small2,small3}.

In this paper, we use the data from the GRB Monitor (GBM) on board Fermi to calibrate the photon (representing electrons) energy content of the GRB jet.
Subsequently, we compare this with the upper limit on proton (turned pion) energy content, given the non-detection of GRB neutrinos.
Furthermore, the first catalog of the Large Area Telescope (LAT) on board Fermi includes 35 GRBs with gamma ray emission above 100  GeV \citep{Ackermann}.
Several models have been proposed to explain this high energy emission \citep{meszaAFTERGLOW, gg03, gpw11}  including hadronic models \citep{GupZhang, bd2000}.
The same photon-hadron process that produces the charged pions and subsequently the 100 TeV neutrinos, would also generate neutral pions that decay to photons of similar energy.
These high-energy photons have been hypothesized to cascade through pair production processes down to the GeV regime, where they can escape the jet and be observed by LAT.
Within this scenario, we use the observed GeV burst fluence to put another upper limit on the energy content of the protons in the jet.



\section{Methodology}
In a relativistic outflow (fireball), the energy carried by the hadrons can be dissipated internally, or through interactions with ambient matter.
Thus, a substantial part of the bulk kinetic energy is converted to internal energy, which is then distributed between electrons, protons and the magnetic field.
We denote the ratio between the energy carried by electrons and that of the protons as $f_e $. 
If the plasma is in equipartition, $f_e \approx 1$, but in our analysis this is not a requirement.
The internally accelerated electrons presumably are responsible for the keV-MeV photons observed in the GRB, which are emitted through synchrotron or inverse Compton processes.
The measured GBM burst fluence is, thus, proportional to the energy carried by electrons.

Accelerated protons may interact with these ($\sim$ MeV) photons to produce pions via the Delta resonance,

\begin{equation} \label{eq_PG}
     p+\gamma \to \Delta ^{+} \to \bigg\{
    \begin{gathered}
        n+\pi ^{+} \\
        p+\pi ^{0} \\
    \end{gathered}
\end{equation}

\noindent
The branching ratios for $\pi ^{+}$ and $\pi ^{0}$ production in this process are 1/3 and 2/3, respectively.
Taking into account higher-energy resonances, this interaction could lead to a higher yield of charged pions \citep{WB97, Hummer2010}.
The $\pi ^{0}$ decays to two photons, which are discussed below in Sec. \ref{Sec:pi0_estimate}.
The associated proton may interact with the photons to produce secondary pions, which could increase the expected neutrino flux from the GRB, but we neglect these here.
Including them would only tighten the constraints on the hadronic content that we derive below from the non-detection of neutrinos.
The charged pion decays to produce $e$ and $\mu$ neutrinos and anti-neutrinos:

\begin{equation}
\label{eq_pi_decay}
\pi ^{+}\to e^{+}+\nu_{e}+\nu _{\mu }+\bar{\nu }_{\mu }
\end{equation}

\noindent The energy in this decay is split about evenly between the products, i.e. 3/4 of the $\pi ^{+}$ energy goes to neutrinos.

\subsection {Neutrino Fluence Estimate}
\label{Sec:N_estimate}

Although IceCube can detect neutrinos of all flavors \citep{halzen2010}, it is most sensitive to tracks produced by $\nu _\mu$.
On the other hand, the atmospheric background of $\nu _\mu$ is very high, which calls for the exploitation of shower events of $\nu _e$ and $\nu _\tau$ \citep{halzen2010, Aartsen_Sci}.
However, within the short time window of an individual GRB, IceCube is essentially background-free.
In other words, a muon track of sufficiently high energy that is associated in direction and in time with a GRB would most surely be a real detection.
This allows IceCube to exploit its high effective area for $\nu _\mu$ events, without the downside of the $\nu _\mu$ background.
Allowing for effective neutrino oscillations, which results in an equal flux of the three flavors at the detector, and since  in point source searches the detector is about 10 times more sensitive to $\nu _\mu$ than to the other flavors, IceCube can be expected to detect 1/3 of the neutrinos, and hence 1/4 of the $\pi ^{+}$ energy in the form of $\nu _\mu$.

Denoting the fraction of proton energy that goes into pions as $f_{\pi}$,
the fraction of proton energy that ends up in neutrinos is consequently $f_{\pi}/4$.
Simulations by \citet{dafneRATE} suggest that $f_{\pi}\approx 0.2$.
It has been suggested that if the resulting neutron (equation~\ref{eq_PG}) remains in the plasma, $f_{\pi}$ can be much higher \citep{Baerwald2013}.
However, in this work we leave $f_{\pi}$ as a free parameter and attempt to constrain it from the IceCube results.
One can now relate the hadron energy content of the GRB jet to that of the gamma ray emitting electrons through the ratio $f_\pi / f_e$.

The GRB fluence serves as a proxy of the electron energy in the jet.
The electrons are assumed to follow a power-law energy distribution with a spectral slope of $\alpha$, namely $dN_e / dE_e\propto E_e^{-\alpha}$.
Noting that the prompt GRB energy fluence measured in the MeV (GBM) band, $F_\text{GBM}$, is due only to the electron population in a limited energy range $E_{e,\text{min}} - E_{e, \text{max}}$, one can write:

\begin{equation} \label{eq_slope}
F_\text{GBM} \propto \int_{E_{e,\text{min}}}^{E_{e,\text{max}}}{E_e^{1-\alpha}}dE_e
\end{equation}

\noindent Assuming the electrons, protons, and thus (pions and) neutrinos all adhere to the same slope, and that all their energies range the same number of decades, the $\nu _\mu$ energy fluence of a GRB, $F_\nu$, can be directly related to $F_\text{GBM}$ through:

\begin{equation} \label{eq_N_general}
  F_{\nu}=\frac{1}{12}\frac{f_{\pi }}{f_e} \frac{2-\alpha}{E_{e,\text{max}}^{2-\alpha}-E_{e,\text{min}}^{2-\alpha}}F_\text{GBM}
\end{equation}

What is the neutrino spectral slope?
$\alpha = 2$ is motivated by Fermi acceleration, as well as by the IceCube measured slope for diffuse neutrinos \citep{Aartsen_Sci}.
An independent indication of the slope could come from the photon (GRB) spectra.
However, the relation between the photon and neutrino spectral slopes is not totally clear.
While \citet{WB97} assume they are the same, more detailed fireball models use more parameters to relate the two \citep{dafnebig, Becker2010}.
Since we wish to avoid model-dependent assessments, we limit ourselves to the generic slope of $\alpha = 2$.
In Section \ref{slopes}, we will discuss the scenario of a neutrino power-law that deviates from the canonical --2 slope.

In the limit of $\alpha = 2$, equation~(\ref{eq_N_general}) provides a simple expression for
$F_\nu$:

\begin{equation} \label{eq_Fn}
    F_{\nu }
    =\frac{1}{12}\frac{f_{\pi }}{f_e} \frac{F_\text{GBM}}{\ln 10}
\end{equation}

\noindent which is conveniently independent of $E_{e,\text{max}}, E_{e,\text{min}} $.
The factor of $\ln 10$ is due to the fact that the GBM band is roughly two decades of photon energy from 0.01~MeV to 1~MeV, which arises from only one decade of electrons energy (i.e., $E_{e,\text{max}} / E_{e,\text{min}}  =10$), since the energy of photons emitted by both synchrotron and inverse Compton scale as $E_e^2$.
$F_\text{GBM}/(f_e \ln 10)$ in equation (\ref{eq_Fn}) merely represents the total energy in protons.
Using equation (\ref{eq_Fn}), the neutrino number fluence is:

\begin{equation} \label{eq_dNdE}
    \frac{dN_{\nu}}{dE_{\nu}dA}=
     \frac{1}{12}\frac{f_{\pi }}{f_e} \frac{F_\text{GBM}}{\ln 10}E_{\nu}^{-2}
\end{equation}

Employing the IceCube effective area curves $A_\text{eff}(E_\nu)$ as a function of declination, we can now estimate the number $N_{\nu}$ of neutrinos expected to be detected by IceCube for each individual GRB, and as a function of the single parameter $f_{\pi} / f_e$ representing the hadronic fraction in the GRB jet:

\begin{equation} \label{eq_N}
    N_{\nu}=\frac{1}{12}\frac{f_{\pi }}{f_e} \frac{F_\text{GBM}}{\ln 10}\int{A_\text{eff}(E_\nu)E_{\nu}^{-2}}dE_{\nu}
\end{equation}

\begin{figure}
\includegraphics[width=0.5\textwidth]{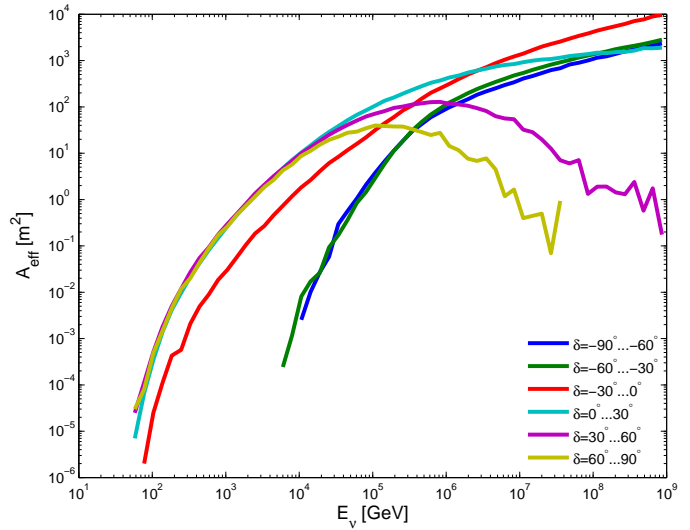}\\
\caption{Effective area of the complete IC86 for $\nu _\mu$ point sources versus neutrino energy (A. Karle, J. Feintzeig, private communications). Different declinations ($\delta$) on the sky are plotted separately.}
\label{EffArea}
\end{figure}

\noindent In Figure \ref{EffArea}, we plot the effective area curves of the complete IC86 array for point source detection of $\nu _\mu$ (A. Karle, J. Feintzeig, private communications).
The efficiency of detecting $\nu _e$ and $\nu _\tau$ from a point source
is much smaller, and we neglect it here.

\begin{figure}
\centering
\includegraphics[width=0.5\textwidth]{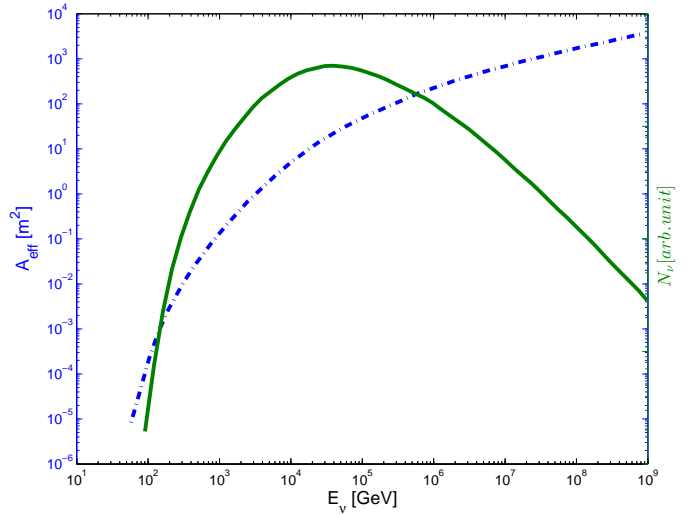}\\
\caption{Effective area (dashed curve) of the complete IC86 for  $\nu _\mu$ point sources versus neutrino energy averaged over declination (see Fig.~\ref{EffArea}) and the expected energy distribution of detected neutrinos given an $E_\nu ^{-2}$ source spectrum (solid curve).}
\label{weighted_Aeff}
\end{figure}

The IceCube effective area curves for negative declinations (overhead at the South Pole) continue to rise with neutrino energy. However, given the declining energy spectrum (equation \ref{eq_dNdE}), the most GRB neutrinos are expected in IceCube around 30 TeV.
This is demonstrated in Figure~\ref{weighted_Aeff}, where the IceCube effective area is averaged over declination and the expected number of neutrinos per logarithmic energy bin is plotted, assuming an $E_\nu ^{-2}$ spectrum.
Spectra that are markedly different could produce neutrinos at much higher energies \citep[e.g.,][]{Razzaque13}.
IceCube is even more sensitive to those, if they exist.

Finally, in addition to the estimate for each GRB, we can obtain the total number of neutrinos expected from the full sample of GBM GRBs, by using their co-added fluences in equation (\ref{eq_N}).

\subsection {GeV Photon Fluence Estimate}
\label{Sec:pi0_estimate}
The photons resulting from $\pi ^{0}$ decay, according to some models, cascade through pair production processes down to GeV energies at which point they can escape the jet.
These photons can be detected by LAT, whose sensitivity ranges from 0.02 - 300 GeV \citep{Atwood}. They would add to any other electron emission at these energies if present. Recall that 2/3 of pions produced by the photon-hadron interactions (equation~\ref{eq_PG}) are $\pi ^{0}$. For a spectral slope of $\alpha = 2$, and by analogy to equation~(\ref{eq_Fn}), their (maximal) expected contribution to the fluence measured by LAT would be:



\begin{equation} \label{eq_Fpi0}
    F_\gamma^{\pi ^0}=\frac{2}{3} \frac{f_{\pi
    }}{f_e} \frac{F_{GBM}}{\ln 10}
\end{equation}

\noindent
If the spectral slope $\alpha$ deviates from 2, the more general factor needs to be used here (c.f., equations~\ref{eq_N_general}) instead of $\ln 10$.
The cascade products would, thus, carry out from the GRB most of the energy that was initially in protons. Additional GeV photon fluence $F_\gamma ^{\pi ^+}$ could arise from the positrons produced in the charged pion decay (equation \ref{eq_pi_decay}) that takes about 1/4 of the $\pi ^{+}$ energy, or 1/12 of the total pion energy

\begin{equation} \label{eq_Fpip}
    F_\gamma^{\pi ^+}=\frac{1}{12} \frac{f_{\pi
    }}{f_e} \frac{F_{GBM}}{\ln 10}
\end{equation}

Finally, the original electrons in the jet may also emit GeV photons with fluence $F_\gamma ^e$.
Accounting for all of these potential GeV photon emission processes, one can express, most generally, the total LAT fluence as:

\begin{equation} \label{eq_Flat}
    F_\text{LAT} = F_\gamma ^{\pi ^0} + F_\gamma ^{\pi ^+} + F_\gamma ^e = \frac{3}{4} \frac{f_{\pi }}{f_e} \frac{F_{GBM}}{\ln 10} +  F_\gamma ^e
\end{equation}

\noindent The hadronic contribution of the detected LAT fluence is the first term on the right hand side of equation (\ref{eq_Flat}), which allows us to write the hadronic fraction of the LAT fluence as:

\begin{equation} \label{eq_fhad}
    f_\text{Had} = \frac{3}{4 \ln 10} \frac{f_{\pi }}{f_e} \frac{F_\text{GBM}}{F_\text{LAT}}
\end{equation}

\noindent
The actual contribution of the hadrons to the GeV emission is model dependent.
However, an absolute upper limit to this contribution is $f_\text{Had} \le 1$.
Since both $F_\text{GBM}$ and $F_\text{LAT}$ are measured quantities, equation (\ref{eq_fhad}) provides an absolute upper limit to $f_\pi / f_e$.

\begin{figure}
  \centering
\includegraphics[width=0.5\textwidth]{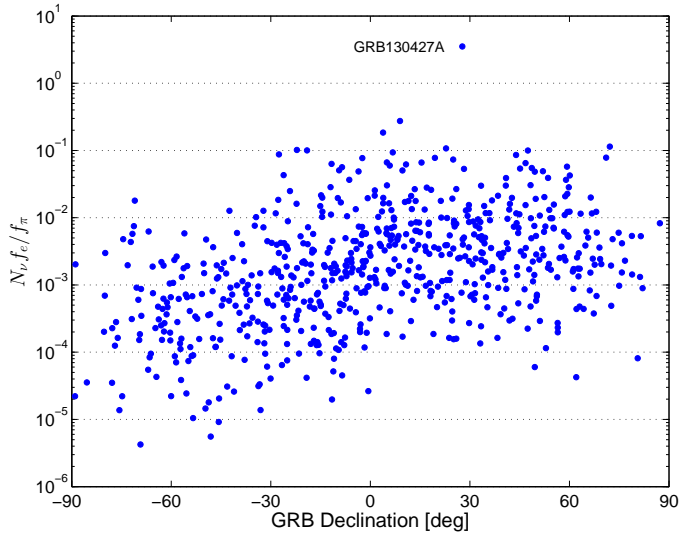}\\
  \caption{Expected number of $\nu _\mu$ from all of the GBM detected GRBs between 2011, Jun and 2014, Feb, as a function of declination, and factored by the unknown electron to pion energy ratio $f_e / f_\pi$. Note the reduced $\nu _\mu$ numbers at low southern declinations where the effective area for TeV energies is smallest (Fig.~\ref{weighted_Aeff}).}
\label{NuNum}
\end{figure}
\

\
\section{Results}
\subsection{Constraints from IceCube Non-Detection}
\label{sec:N_results}

We use equation (\ref{eq_N}) to estimate the expected number of neutrinos from each GRB based on its declination.
For this, we take the fluence of all 668 GRBs between 2011 Jun - 2014 May, from the GBM burst catalog \footnote{http://heasarc.gsfc.nasa.gov/W3Browse/fermi/fermigbrst.html}, co-temporal with the running of IC86.
For simplicity, at this point, we assume the neutrino spectrum has a standard spectral slope of --2, e.g. the neutrino spectrum is that of equation (\ref{eq_dNdE}).
The expected number of IceCube $\nu _\mu$ events factored by the electron to pion energy ratio in the jet, namely $N_\nu (f_e / f_\pi)$, is plotted in Figure \ref{NuNum} versus declination.
Note the exceptionally bright GRB130427A, whose favorable declination of 27.7$^\text{o}$, yields 3.5 expected $\nu _\mu$ neutrinos for $f_e /f_\pi$ of unity, which is an order of magnitude more than any other GRB.
The lack of detected neutrinos from GRB130427A was recently discussed by \citet{gao13} in the context of the physical jet parameters.

The IceCube non-detection of neutrinos during the GRB130427A time window
constrains the ratio of $f_\pi / f_e$.
Using Bayesian statistics, \citet{Astone} propose that with no background, a non-detection implies an interval of $0 - N$ neutrinos at a confidence-level-equivalent(CL) of $1 - e^{-N}$.
For a 95\% CL, $N = 3$.
Consequently, GRB130427A alone implies $N_\nu (f_e / f_\pi) < 3 (f_e / f_\pi) $ at 95\% CL. Or, $f_\pi / f_e \lesssim 0.85$ at 95\% CL, and $f_\pi / f_e \lesssim 0.35$ at 68\% CL.
This limit can be tightened, if we take into account that no neutrinos were detected from any GRB in Figure~\ref{NuNum}. The total expected $\nu _\mu$ counts over the 36 months period from all 668 GRBs combined increases the expected counts by (only) 146\% over that of GRB130427A, i.e. $N_\nu (f_e /f_\pi) = 8.7$, leading to a stricter constraint of $f_\pi / f_e < 0.35\,(0.12)$ at 95\%\,(68\%) CL.

A reasonable next step would be to include in the above analysis also the non-detection of GRB neutrinos since the IceCube 2008-2009 season \citep{icecubenature}.
This exercise is somewhat complicated by the gradual increase in number of strings (i.e., effective area), and varying operation periods in each season.
Nonetheless, the operation-period weighted sum of the IceCube seasonal effective area, based on \citet{Aartsen_Apj}, suggests that the five seasons before IceCube's completion are roughly equivalent to 788 days or $\sim$26 months of operation with 86 strings. The estimated IC86 equivalent periods are listed in Table \ref{tab1}. We now assume that the GRB occurrence over these years is the same as during the IC86 36 months sample, but excluding the anomalously bright GRB130427A, i.e., $N_\nu = 5.2(f_e/ f_\pi$) (c.f., 8.7 with GRB130427A) $\nu _\mu$ events are anticipated over 36 months. Thus, for 26 months we anticipate $N_\nu (f_e / f_\pi) = 3.7$ (to be added to the 8.7). All in all, the constraint on the hadronic component in GRB jets over the operation period of IceCube since 2008 improves to $f_\pi / f_e < 3 / 12.4 \approx 0.24\,(0.08)$ at 95\%\,(68\%) CL. Note that for a branching ratio more favorable of charged pions in equation~\ref{eq_PG}, e.g. 1/2 and 1/2 \citep{WB97} instead of 1/3 and 2/3, all of these constraints would be stronger, reducing $f_\pi / f_e$ by a factor of 3/2.

\begin{table}
    \begin{tabular}{l c c c c}
    \hline \hline
    Season & \# Strings & $A_{eff}$ & Period  & IC86 equivalent \\
     &  &  ($\%$ of IC86) & (days) & (days) \\ \hline
    2011-2014 & 86 & 100  & 1096 & 1096 \\
    2010-2011 & 79 & 90  & 316 & 284.4 \\
    2009-2010 & 59 & 80  & 348 & 278.4 \\
    2008-2009 & 40 & 60  & 375 & 225   \\
            Total  &    &     && 1884 \\ \hline
    \end{tabular}

    \caption{IceCube seasons}
    \label{tab1}
\end{table}

\subsection{Constraints from LAT Fluence}
\label{sec:LAT_results}

The most conservative estimation for the hadronic contribution to the GeV photon fluence measured by LAT is $f_\text{Had} \le 1$, i.e. all LAT fluence is hadronic (via  pair-photon cascades).
Using the typical ratio $F_\text{GBM} / F_\text{LAT} \approx 10$ \citep{Ackermann} in equation (\ref{eq_fhad}) we can constrain the typical GRB hadronic fraction to be $f_\pi / f_e \lesssim 0.3$.
For LAT detected bursts, as with the neutrinos, this analysis can be carried out for each individual GRB.
In the extreme cases, where $F_\text{GBM} / F_\text{LAT}$ is lowest, of the order of unity, for example GRB\,090510 and GRB\,080916C \citep{Ackermann}, the constrain on the hadronic fraction is strongest, i.e., $f_\pi / f_e \lesssim 0.03$.
Specifically for the bright GRB130427A $F_\text{GBM} / F_\text{LAT} \approx 5$, which yields $f_\pi / f_e \lesssim 0.15$.
For a branching ratio less favorable for $\pi ^0$ in equation~\ref{eq_PG}, e.g. 1/2 and 1/2 \citep{WB97} instead of 1/3 and 2/3, this constrain would be weaker, increasing all the above values of $f_\pi / f_e$ by a factor of 3/2.

Note that the constraint from LAT on $f_\pi / f_e  \lesssim 0.15$ for GRB\,130427A appears to be tighter than that from the non-detection of neutrinos ($f_\pi / f_e \lesssim 0.85$, Sec. \ref{sec:N_results}).
However, the LAT constrain relies on the assumption that all $\pi ^0$ energy cascades down to GeV photons,
while the neutrino constraint assumes nothing but that charged pions are produced in the jet.

\subsection{Dependence on Spectral Slope}
\label{slopes}
In this section, we analyze the sensitivity of the number of GRB neutrinos $N_\nu (f_e / f_{\pi})$ expected from the analysis in Sec.~\ref{sec:N_results} to the assumed neutrino spectral slope $\alpha$.
Therefore, we repeat the analysis for the number of neutrinos expected from GRB\,130427A with varying spectral slopes.
Instead of the simple expression for $N_\nu$ obtained in equation~(\ref{eq_Fn}), we need to use the more general form of equation~(\ref{eq_slope}).
As before, we assume all particles (protons, electrons, neutrinos) have the same spectral slope.
Equation (\ref{eq_N}) thus obtains the more general form of:

\begin{equation} \label{eq_Nslope}
  N_{\nu}=\frac{1}{4}\frac{f_{\pi }}{f_e} \frac{(2-\alpha)}{E_{e,\text{max}}^{2-\alpha}-E_{e,\text{min}}^{2-\alpha}}F_\text{GBM}\int{A_\text{eff}(E_{\nu})E_{\nu}^{-\alpha}}dE_{\nu}
\end{equation}

\begin{figure}
  \centering
\includegraphics[width=0.45\textwidth]{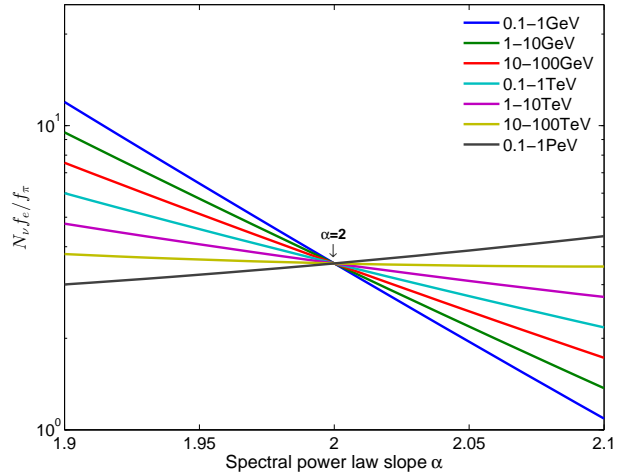}\\
  \caption{Expected number of nuetrinos from GRB\,130427A as a function of the spectral slope $\alpha$, factored by the unknown electron to pion energy ratio $f_e / f_\pi$, and for a range of gamma-ray emitting electron energies as represented by the different colors.} \label{Alpha}
\end{figure}

The number of expected neutrinos in Eq.~\ref{eq_Nslope} now depends on $E_{e,\text{max}}$ and $E_{e,\text{min}}$, as opposed to the special case  of $\alpha = 2$ (c.f., equation~\ref{eq_N}).
We retain the single-decade electron energy window $E_{e,\text{max}} / E_{e,\text{min}} = 10$, which still corresponds to the two decades of photon energy detected by GBM (Sec.~\ref{Sec:N_estimate}).
It is less clear what should be assumed for $E_{e,\text{max}}$ and $E_{e,\text{min}} $, as the energy of electrons producing the GBM (MeV) photons strongly depends on the mechanism, whether synchrotron or inverse Compton, and on the energy density of the magnetic field or seed photons, respectively.
We therefore test a range of electron energies for slope parameters $1.9<\alpha <2.1$.
The results are plotted in Fig.~\ref{Alpha}.
It can be seen that the expected number of neutrinos $N_\nu (f_e / f_{\pi})$ is rather sensitive to the spectral slope, and  
can vary by up to an order of magnitude for $1.9<\alpha <2.1$ (Figure \ref{Alpha}), especially for low electron energies of $\sim$ GeV. 
If the electron energies are much higher, namely TeV and above, the constrain on  $f_\pi / f_e$ would depend more weakly on the assumed spectral slope.

From Fig.~\ref{Alpha}, it is clear that the expected number of neutrinos can vary dramatically even if the spectral slope is only slightly different than $\alpha = 2$. Since there could be a distribution of neutrino slopes around $\alpha = 2$, some of them would necessarily produce appreciably higher numbers than the values plotted in Fig.~\ref{NuNum}. In that sense, our estimates above are the most conservative. 

\begin{figure}
  \centering
\includegraphics[width=0.45\textwidth]{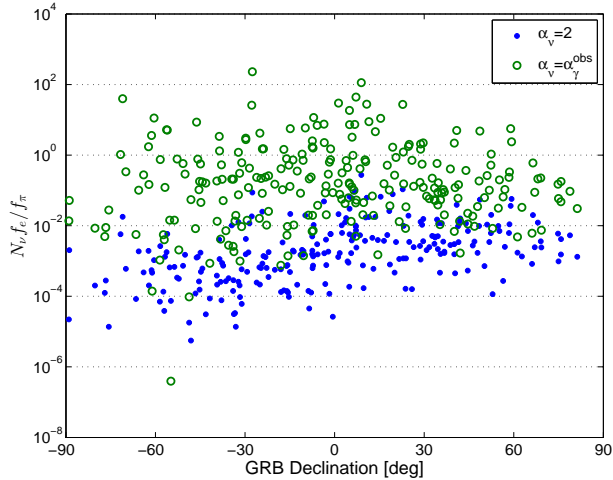}\\
  \caption{Expected number of $\nu _\mu$ from 250 GBM detected GRBs, which have a measured spectral slope, as a function of declination, and factored by the unknown electron to pion energy ratio $f_e / f_\pi$. The neutrino spectral slope is assumed to be that of the photons.} \label{slope}
\end{figure}

In order to further demonstrate the dependence on $\alpha$, we computed $N_\nu (f_e / f_{\pi})$ for the 250 individual bursts whose photon spectral index is given in the GBM catalog. 
For this purpose, we assume that the neutrino spectral index is the power law spectral index of the GRB \citep{WB99}.
Since on average, the measured slope in this sample is $<\alpha > \approx 1.5$, the expected number of neutrinos increases by orders of magnitude compared to our previous estimate.
For example, if the GRBs are emitted by electrons in the 10 -- 100 GeV range \citep[][see Fig.~\ref{Alpha} above]{Bosnjak09},
$N_\nu (f_e / f_{\pi})$ in this limited sample increases from $\sim$2 to $\sim$770.
The results for each individual GRB are plotted in Fig.~\ref{slope}, which shows many GRBs yielding high values of $N_\nu (f_e / f_{\pi})$, and thus tightening the upper limit on the $f_\pi / f_e$  by more than two orders of magnitude.

The precise neutrino numbers here depend strongly on the prescription for assigning a spectral slope to the neutrinos and on the chosen electron energy range, while the previous estimate with $\alpha = 2$ depends only on the total GRB fluence $F_\text{GBM}$.
Most importantly, any distribution of neutrino slopes around $\alpha = 2$ would result in a huge increase in the number of predicted neutrinos, and make the contrast with the non-detection even starker.

\section{Conclusions}
We find that both the non-detection of any neutrinos from GRBs and the observed GeV GRB fluence point consistently to a low hadronic energy fraction in the GRB jet.
More quantitatively, the lack of detected neutrinos from Fermi/GBM GRBs since 2008 points to $f_\pi / f_e \lesssim 0.24$ with a 95\% CL.
As far as we know, this is the first time this fraction has been constrained from observations.
These numbers hold for a canonical spectral slope of $F_\nu \propto E_\nu ^{-2}$, which is expected from standard acceleration mechanisms and is consistent with the observed spectrum of diffuse neutrinos reported recently by the IceCube collaboration.
Given that there could be a distribution of neutrino spectral slopes, the constrain on $f_\pi / f_e$ would tighten, even by a few orders of magnitude.

The obtained value of $f_\pi / f_e \lesssim 0.2$ is still consistent with the values of $f_e\approx 1$ \citep{WB97}, and of $f_\pi \approx 0.2$ \citep{dafneRATE}.
More realistic models, however, that include the cooling of electrons in the GRB jet predict $f_e \ll 1$ \citep[e.g.,][]{gao13}.
Given the limited efficiency of pion production, these models are in strong contrast with the current findings.

The observed LAT fluence from GRBs, independent of neutrino physics, provides on average a constrain of $f_\pi / f_e \lesssim 0.3$, which is consistent with the neutrino estimate. However, much more stringent limits of $f_\pi / f_e \lesssim 0.03$ are obtained for individual GRBs whose MeV (GBM) to GeV (LAT) fluence ratio is particularly low ($\lesssim 1$).

The presently found low hadronic fractions,  along with the failure of GRBs to explain the observed UHECRs \citep{icecubenature} contribute to the growing questions regarding the physical presence of PeV protons in GRBs.
Ultimately, these protons would necessarily produce neutrinos that would need to be observed by IceCube.
The longer IceCube goes without detecting a GRB neutrino, the constraint on $f_ \pi / f_e$ will tighten.

\acknowledgments
We thank Albrecht Karle and Jacob Feintzeig for providing the unpublished effective area curves and for most helpful discussions regarding the response of the IceCube detector.
We acknowledge useful comments on the manuscript from Charles Dermer, Hagar Landsman, Francis Halzen, Arnon Dar, and Demos Kazanas.
This research is supported by the I-CORE program of the Planning and Budgeting Committee and the Israel Science Foundation (grant numbers 1937/12 and 1163/10), and by a grant from Israel's Ministry of Science and Technology.
D.G. is supported by a grant from the U.S. Israel Binational Science Foundation.

\end{document}